\documentclass[journal,11pt,draftclsnofoot,onecolumn]{IEEEtran}

\IEEEoverridecommandlockouts                              

\usepackage{graphicx}
\usepackage[sort,compress]{cite}
\usepackage{amsthm,amssymb,amscd,amssymb,amsmath,mathrsfs}
\usepackage{epstopdf}
\usepackage{balance}
\usepackage[noend]{algorithm} 
\usepackage{algorithmicx} 
\usepackage[noend]{algpseudocode}
\usepackage{multirow} 
\usepackage{amsmath}
\usepackage{xcolor}
\usepackage{lineno}

\makeatletter

\newcommand{\Rmnum}[1]{\expandafter\@slowromancap\romannumeral #1@}
\makeatother

\theoremstyle{plain}
\newtheorem{theorem}{Theorem}
\newtheorem{prop}{Proposition}
\newtheorem{lem}{Lemma}

\theoremstyle{definition}

\theoremstyle{remark}

\begin{document}
%
\title{Characterizations and Effective Computation of Supremal Relatively Observable Sublanguages*}

\author{Kai Cai$^{1}$ and Renyuan Zhang$^{2}$, W.M. Wonham$^{3}$
\thanks{*This work was supported in part by JSPS KAKENHI Grant no. JP16K18122 and Program to Disseminate Tenure Tracking System, MEXT, Japan;
the National Nature Science Foundation, China, Grant no. 61403308;
the Natural Sciences and Engineering Research Council, Canada, Grant no. 7399.
}
\thanks{$^{1}$K. Cai is with Urban Research Plaza, Osaka City University, Japan
        ({\tt\small kai.cai@eng.osaka-cu.ac.jp})}%
\thanks{$^{2}$R. Zhang is with School of Automation, Northwestern Polytechnical University, China
        ({\tt\small ryzhang@nwpu.edu.cn})}%
\thanks{$^{3}$W.M. Wonham is with the Systems Control Group, Department of Electrical and
        Computer Engineering, University of Toronto, Canada
        ({\tt\small wonham@control.utoronto.ca}).}%
}

\maketitle


\thispagestyle{empty} \pagestyle{plain}

\begin{abstract}
Recently we proposed \emph{relative observability} for supervisory
control of discrete-event systems under partial observation.
Relative observability is closed under set unions and hence there
exists the supremal relatively observable sublanguage of a given
language. In this paper we present a new characterization of
relative observability, based on which an operator on languages is proposed whose largest fixpoint is the supremal relatively observable sublanguage.
Iteratively applying this operator yields a monotone sequence of languages; exploiting the linguistic concept of \emph{support}  based on Nerode equivalence, we prove for regular languages that the sequence converges finitely to the supremal relatively observable sublanguage, and the operator is effectively computable. Moreover, for the purpose of control, we propose a second operator that in the regular case computes the supremal relatively observable and controllable sublanguage. The computational effectiveness of the operator is demonstrated on a case study.
\end{abstract}



\section{Introduction} \label{sec1_intro}

In \cite{CaiZhaWon_TAC14} we proposed {\it relative observability}
for supervisory control of discrete-event systems (DES) under
partial observation. The essence of relative observability is to set
a fixed ambient language relative to which the standard
observability conditions \cite{LinWon:88Obs} are tested. Relative
observability is proved to be stronger than observability
\cite{LinWon:88Obs,CDFV:88}, weaker than normality
\cite{LinWon:88Obs,CDFV:88}, and closed under arbitrary set unions.
Therefore the supremal relatively observable sublanguage of a given
language exists, and we developed an automaton-based algorithm to compute the supremal sublanguage.

In this paper and its conference precursor \cite{CaiWonham:WODES16}, 
we present a new characterization of relative observability. 
The original definition of relative observability in
\cite{CaiZhaWon_TAC14} was formulated in terms of \emph{strings},
while the new characterization is given in \emph{languages}. 
Based on this characterization, we propose an operator on languages, whose largest fixpoint is precisely the supremal relatively observable sublanguage. 
Iteratively applying this operator yields a monotone sequence of languages. In the case where the relevant languages are regular, we prove that the sequence converges finitely to the supremal relatively observable sublanguage, and the operator is effectively computable.

This new computation scheme for the supremal sublanguage 
is given entirely in terms of languages, and
the convergence proof systematically exploits the concept of \emph{support} (\cite[Section~2.8]{SCDES})
based on Nerode equivalence relations \cite{HopUll:79}. The solution therefore separates out
the linguistic essence of the problem from the implementational
aspects of state computation using automaton models.  This
approach is in the same spirit as \cite{WonRam:87} for
controllability, namely operator fixpoint and successive
approximation.

Moreover, the proposed language-based scheme allows more straightforward implementation, 
as compared to the automaton-based algorithm in \cite{CaiZhaWon_TAC14}. In particular, we show that the language operator used in each iteration of the language-based scheme may be decomposed into a series of standard or well-known language operations (e.g. complement, union, subset construction); therefore off-the-shelf algorithms may be suitably assembled to implement the computation scheme. On the other hand, both the language and automaton-based algorithms have (at least) exponential complexity in the worst case, which is the unfortunate nature of supervisor synthesis under partial observation. 
Our previous experience with the automaton-based algorithm in
\cite{CaiZhaWon_TAC14} suggests that computing the supremal
relatively observable sublanguage is fairly delicate and thus prone
to error.  Hence, it is advantageous to have two algorithms at hand so that one can double check the computation results,
thereby ensuring presumed correctness based on consistency. 

Finally, for the purpose of supervisory control under partial observation, we combine relative observability with controllability. In particular, we propose an operator which in the regular case effectively computes the supremal relatively observable and controllable sublanguage. We have implemented this operator and tested its effectiveness on a case study.

The rest of the paper is organized as follows. In
Section~\ref{sec2_relobs} we present a new characterization of
relative observability, and an operator on languages that yields an iterative scheme to compute the supremal relatively observable sublanguage. 
In Section~\ref{sec3_reglan} we prove that in the case of regular languages, the iterative scheme generates a monotone sequence of languages that is finitely convergent to
the supremal relatively observable sublanguage.
In Section~\ref{sec4_obscon} we combine relative observability and controllability, and propose an operator that effectively computes the supremal relatively observable and controllable sublanguage. 
Section~\ref{sec5_examp} presents illustrative examples, and finally in Section~\ref{sec6_concl} we state conclusions.

This paper extends its conference precursor \cite{CaiWonham:WODES16} in the following respects. (1) In the main result of Section~\ref{sec3_reglan}, Theorem~\ref{thm:convergence}, the bound on the size of the supremal sublanguage is tightened and the corresponding proof given. (2) The effective computability of the proposed operator is shown in Subsection~\ref{subsec:effcom}. (3) Relative observability is combined with controllability in Section~\ref{sec4_obscon}, and a new operator is presented that effectively computes the supremal relatively observable and controllable sublanguage. (4) A case study is given in Subsection~\ref{subsec:agv} to demonstrate the effectiveness of the newly proposed computation schemes.


\section{Characterizations of Relative Observability and its Supremal Element}
\label{sec2_relobs}

In this section, the concept of relative observability proposed in
\cite{CaiZhaWon_TAC14} is first reviewed. Then we present a new characterization of 
relative observability, together with a fixpoint characterization of the supremal relatively observable sublanguage.

\subsection{Relative Observability}

Let $\Sigma$ be a finite event set.
A string $s \in \Sigma^*$ is a \emph{prefix} of another string $t \in \Sigma^*$, written $s \leq
t$, if there exists $u \in \Sigma^*$ such that $s u = t$.
Let $L \subseteq \Sigma^*$ be a language.
The \emph{(prefix) closure} of $L$ is
$\overline{L} := \{ s \in \Sigma^* \ |\ (\exists t
\in L)\ s \leq t \}$.
For partial observation, let the event set $\Sigma$ be partitioned
into $\Sigma_o$, the observable event subset, and $\Sigma_{uo}$, the
unobservable subset (i.e. $\Sigma = \Sigma_o \dot{\cup}
\Sigma_{uo}$). Bring in the \emph{natural projection} $P : \Sigma^*
\rightarrow \Sigma_o^*$ defined according to
\begin{equation} \label{eq:natpro}
\begin{split}
P(\epsilon) &= \epsilon, \ \ \epsilon \mbox{ is the empty string;} \\
P(\sigma) &= \left\{
  \begin{array}{ll}
    \epsilon, & \hbox{if $\sigma \notin \Sigma_o$,} \\
    \sigma, & \hbox{if $\sigma \in \Sigma_o$;}
  \end{array}
\right.\\
P(s\sigma) &= P(s)P(\sigma),\ \ s \in \Sigma^*, \sigma \in \Sigma.
\end{split}
\end{equation}
In the usual way, $P$ is extended to $P : Pwr(\Sigma^*) \rightarrow
Pwr(\Sigma^*_o)$, where $Pwr(\cdot)$ denotes powerset. Write $P^{-1}
: Pwr(\Sigma^*_o) \rightarrow Pwr(\Sigma^*)$ for the
\emph{inverse-image function} of $P$.


Throughout the paper, let $M$ denote the marked behavior of the plant to be controlled,
and $C \subseteq M$ an imposed specification language.
Let $K \subseteq C$. We say
that $K$ is \emph{relatively observable} (with respect to $M$,
$C$, and $P$), or simply $C$-observable, if the following
two conditions hold:
\begin{align*}
&\mbox{(i)}\ (\forall s,s' \in \Sigma^*, \forall \sigma \in
\Sigma)\ s\sigma \in \overline{K}, s' \in
\overline{C}, s'\sigma \in \overline{M}, P(s) = P(s') \Rightarrow s'\sigma \in \overline{K} \\
&\mbox{(ii)}\ (\forall s,s' \in \Sigma^*)\ s \in K, s' \in
\overline{C} \cap M, P(s) = P(s') \Rightarrow s' \in K.
\end{align*}
In words, relative observability of $K$ requires for every lookalike
pair $(s,s')$ in $\overline{C}$ that (i) $s$ and $s'$ have identical
one-step continuations, if allowed in $\overline{M}$, with respect
to membership in $\overline{K}$; and (ii) if each string is in
$M$ and one actually belongs to $K$, then so does the
other. Note that the tests for relative observability of $K$ are not
limited to the strings in $\overline{K}$ (as with standard
observability \cite{LinWon:88Obs,CDFV:88}), but apply to all
strings in $\overline{C}$; for this reason, one may think of $C$ as the
\emph{ambient} language, relative to which the conditions (i) and (ii) are tested.

We have proved in \cite{CaiZhaWon_TAC14} that in general, relative
observability is stronger than observability, weaker than normality,
and closed under arbitrary set unions. Write
\begin{align} \label{eq:O_C}
\mathcal {O}(C) = \{K \subseteq C \ |\ \mbox{$K$ is $C$-observable }
\}
\end{align}
for the family of all $C$-observable sublanguages of $C$. Then
$\mathcal {O}(C)$ is nonempty (the empty language $\emptyset$
belongs) and contains a unique supremal element
\begin{align} \label{eq:supO}
\sup \mathcal {O}(C) := \bigcup \{ K \ |\ K \in \mathcal {O}(C) \} 
\end{align}
i.e. the supremal relatively observable sublanguage of $C$.

\subsection{Characterization of Relative Observability}

For $N \subseteq \Sigma^*$, write $[N]$ for $P^{-1}P(N)$, namely the
set of all lookalike strings to strings in $N$. A language $N$
is \emph{normal} with respect to $M$ if $[N] \cap M = N$.
For $K \subseteq \Sigma^*$ write
\begin{align} \label{eq:N_N}
\mathcal {N}(K, M) = \{K' \subseteq K \ |\ [K'] \cap M = K'
\}.
\end{align}
Since normality is closed under union, $\mathcal {N}(K,M)$ has a
unique supremal element $\sup \mathcal {N}(K,M)$ which may be
effectively computed \cite{ChoMarcus:89,BrandtLin:90}.

Write
\begin{align} \label{eq:Csigma}
\overline{C}.\sigma := \{s\sigma \ |\ s \in \overline{C}\},\ \
\sigma \in \Sigma.
\end{align}
Let $K \subseteq C$ and define
\begin{align} \label{eq:D}
D(\overline{K}) := \bigcup \left\{ [ \overline{K} \cap
\overline{C}.\sigma ] \cap  \overline{C}.\sigma \ |\ \sigma \in \Sigma \right\}.
\end{align}
Thus $D(\overline{K})$ is the collection of strings in the form $t\sigma$ ($t \in \overline{C}$, $\sigma \in \Sigma$), that are lookalike to the strings in $\overline{K}$ ending with the same event $\sigma$.
Note that if $K = \emptyset$ then $D(\overline{K}) = \emptyset$.
This language $D(\overline{K})$ turns out to be key to the following characterization of relative observability. 

\begin{prop} \label{prop:chara}
Let $K \subseteq C \subseteq M$. Then $K$ is
$C$-observable if and only if
\begin{align*}
& {\rm (i')}\ D(\overline{K}) \cap \overline{M} \subseteq
\overline{K} \\
& {\rm (ii')}\ [K] \cap \left( \overline{C} \cap M
\right) = K.
\end{align*}
\end{prop}

Note that condition (i$'$) is in a form similar to controllability
of $K$ \cite{WonRam:87} (i.e. $\overline{K}\Sigma_{u} \cap \overline{M} \subseteq
\overline{K}$, where $\Sigma_{u}$ is the uncontrollable event set),
although the expression $D(\overline{K})$ appearing here is more complicated owing to
the presence of the normality operator $[\cdot]$. Condition (ii$'$) is normality of $K$ with
respect to $\overline{C} \cap M$.

\emph{Proof of Proposition~\ref{prop:chara}.} We first show that
(i$'$) $\Leftrightarrow$ (i), and then (ii$'$) $\Leftrightarrow$
(ii).

1. (i$'$) $\Rightarrow$ (i). Let $s, s' \in \Sigma^*$, $\sigma \in
\Sigma$, and assume that $s\sigma \in \overline{K}$, $s' \in
\overline{C}$, $s'\sigma \in \overline{M}$, and $P(s) = P(s')$. It
will be shown that $s'\sigma \in \overline{K}$. Since $K \subseteq
C$, we have $\overline{K} \subseteq \overline{C}$ and
\begin{align*}
s\sigma \in \overline{K} &\Rightarrow s\sigma \in \overline{K} \cap \overline{C}.\sigma \\
&\Rightarrow s'\sigma \in [\overline{K} \cap \overline{C}.\sigma] \\
&\Rightarrow s'\sigma \in [\overline{K} \cap \overline{C}.\sigma] \cap \overline{C}.\sigma \\
&\Rightarrow s'\sigma \in D(\overline{K}) \\
&\Rightarrow s'\sigma \in D(\overline{K}) \cap \overline{M} \\
&\Rightarrow s'\sigma \in \overline{K}  \ \ \ (\mbox{by (i$'$)}).
\end{align*}

2. (i$'$) $\Leftarrow$ (i). Let $s \in D(\overline{K}) \cap \overline{M}$.
According to (\ref{eq:D}) $\epsilon \notin D(\overline{K})$;
thus $s \neq \epsilon$.
Let $s = t\sigma$ for some $t\in \Sigma^*$ and $\sigma \in \Sigma$.
Then
\begin{align*}
s \in D(\overline{K}) \cap \overline{M} &\Rightarrow t\sigma \in
[\overline{K} \cap \overline{C}.\sigma] \cap \overline{C}.\sigma \cap \overline{M} \\
&\Rightarrow t \in \overline{C}, t\sigma \in \overline{M}, \\
&\hspace{0.0cm} (\exists t' \in \Sigma^*) (P(t)=P(t'), t'\sigma \in
\overline{K} \cap \overline{C}.\sigma) \\
&\Rightarrow t\sigma \in \overline{K},\ \ \ \mbox{(by (i))}\\
&\Rightarrow s \in \overline{K}.
\end{align*}

3. (ii$'$) $\Rightarrow$ (ii). Let $s, s' \in \Sigma^*$ and assume
that $s \in K$, $s' \in \overline{C} \cap M$, and $P(s) =
P(s')$. Then
\begin{align*}
s \in \overline{K} &\Rightarrow s' \in [\overline{K}] \\
&\Rightarrow s' \in [\overline{K}] \cap \overline{C} \cap M \\
&\Rightarrow s'\sigma \in K \ \ \ (\mbox{by (ii$'$)}).
\end{align*}

4. (ii) $\Rightarrow$ (ii$'$). ($\supseteq$) holds because $K
\subseteq [K]$ and $K \subseteq \overline{C} \cap M$. To
show ($\subseteq$), let $s \in [K]$ and $s \in \overline{C} \cap
M$. Then there exists $s' \in K$ such that $P(s)=P(s')$.
Therefore by (ii) we derive $s \in K$. \hfill $\square$

Thanks to the characterization of relative observability in
Proposition~\ref{prop:chara}, we rewrite $\mathcal {O}(C)$ in
(\ref{eq:O_C}) as follows:
\begin{align} \label{eq:NewOC}
\mathcal {O}(C) = \{K \subseteq C \ |\ D(\overline{K}) \cap \overline{M} \subseteq \overline{K}\ \&\ [K] \cap \left( \overline{C} \cap M \right) = K\}.
\end{align}
In the next subsection, we will characterize the supremal element $\sup \mathcal
{O}(C)$ as the largest fixpoint of a language operator.

\subsection{Fixpoint Characterization of $\sup \mathcal {O}(C)$}

For a string $s \in \Sigma^*$, write $\bar{s}$ for $\overline{\{s\}}$, the set of prefixes of $s$. 
Given a language $K \subseteq \Sigma^*$, let 
\begin{align} \label{eq:FK}
F(K) := \{ s \in \overline{K} \ |\ D(\bar{s}) \cap \overline{M} \subseteq \overline{K} \}.
\end{align}

\begin{lem} \label{lem:FK}
$F(K)$ is closed, i.e. $\overline{F(K)}=F(K)$. Moreover, if $K \in \mathcal {O}(C)$, then $F(K) = \overline{K}$.
\end{lem}

{Proof.} First, let $s \in \overline{F(K)}$; then
there exists $w \in \Sigma^*$ such that $sw \in F(K)$, i.e. $sw \in \overline{K}$ and
$D(\overline{sw}) \cap \overline{M} \subseteq \overline{K}$.
It follows that $s \in \overline{K}$ and $D(\overline{s}) \cap \overline{M} \subseteq \overline{K}$,
namely $s \in F(K)$. This shows that $\overline{F(K)} \subseteq F(K)$; the other direction $\overline{F(K)} \supseteq F(K)$ is automatic. 

Next, suppose that $K \in \mathcal {O}(C)$; by (\ref{eq:NewOC}) we have $D(\overline{K}) \cap \overline{M} \subseteq \overline{K}$. Let $s \in \overline{K}$; it will be shown that $D(\bar{s}) \cap \overline{M} \subseteq \overline{K}$. Taking an arbitrary string $t \in D(\bar{s}) \cap \overline{M}$, we derive
\begin{align*}
& t \in \bigcup \left\{ [ \overline{s} \cap \overline{C}.\sigma ] \cap  \overline{C}.\sigma \ |\ \sigma \in \Sigma \right\} \cap \overline{M} \\
\Rightarrow & t \in \bigcup \left\{ [ \overline{K} \cap \overline{C}.\sigma ] \cap  \overline{C}.\sigma \ |\ \sigma \in \Sigma \right\} \cap \overline{M} \\
\Rightarrow & t \in \overline{K}.
\end{align*}
This shows that $s \in F(K)$ by (\ref{eq:FK}), and hence $\overline{K} \subseteq F(K)$. The other direction $F(K) \supseteq \overline{K}$ is automatic. 
\hfill $\square$

Now define an operator $\Omega : Pwr(\Sigma^*) \rightarrow Pwr(\Sigma^*)$ according to
\begin{align} \label{eq:Omega}
\Omega(K) := \sup \mathcal {N} \big( K \cap F(K),\ \overline{C} \cap M \big),\ \ \ K \in Pwr(\Sigma^*).
\end{align}

A language $K$ such that $K = \Omega(K)$ is called a \emph{fixpoint}
of the operator $\Omega$. The following proposition characterizes $\sup \mathcal {O}(C)$ as the \emph{largest} fixpoint of $\Omega$.

\begin{prop} \label{lem:fixpoint}
$\sup \mathcal {O}(C) = \Omega(\sup \mathcal {O}(C))$, and $\sup
\mathcal {O}(C) \supseteq K$ for every $K$ such that $K =
\Omega(K)$.
\end{prop}

\emph{Proof.} Since $\sup \mathcal {O}(C) \in \mathcal {O}(C)$, we have
\begin{align*}
\Omega(\sup \mathcal {O}(C)) &= \sup \mathcal {N} \big( \sup \mathcal
{O}(C) \cap F(\sup \mathcal {O}(C)), \overline{C} \cap M \big) \\
&= \sup \mathcal {N}(\sup \mathcal {O}(C) \cap \overline{\sup
\mathcal {O}(C)}, \overline{C} \cap M) \\
&= \sup \mathcal {N}(\sup \mathcal {O}(C), \overline{C} \cap M) \\
&= \sup \mathcal {O}(C).
\end{align*}

Next let $K$ be such that $K = \Omega(K)$. To show that $K \subseteq \sup \mathcal {O}(C)$, 
it suffices to show that $K \in \mathcal {O}(C)$. From 
\begin{align*}
K = \Omega(K) := \sup \mathcal {N} \big( K \cap F(K),\ \overline{C} \cap M \big)
\end{align*}
we have $K \subseteq K \cap F(K)$. But $K \cap F(K) \subseteq K$. Hence, in fact, $K = K \cap F(K)$.
This implies that $K = \sup \mathcal {N} \big( K,\ \overline{C} \cap M \big)$; namely $K$ is normal with respect to $\overline{C} \cap M$.

On the other hand, by $K = K \cap F(K) \subseteq F(K)$, we have $\overline{K} \subseteq \overline{F(K)} = F(K)$. But $F(K) \subseteq \overline{K}$ by definition; therefore $\overline{K} = F(K)$. In what follows it will be shown that $D(F(K)) \cap \overline{M} \subseteq F(K)$,
which is equivalent to $D(\overline{K}) \cap \overline{M} \subseteq \overline{K}$.
Let $s \in D(F(K)) \cap \overline{M}$. As in the proof of Proposition~\ref{prop:chara} (item 2),  we know that $s \neq \epsilon$.
So let $s = t\sigma$ for some $t\in \Sigma^*$ and $\sigma \in \Sigma$. Then
\begin{align*}
s \in D(F(K)) \cap \overline{M} &\Rightarrow t\sigma \in
[F(K) \cap \overline{C}.\sigma] \cap \overline{C}.\sigma \cap \overline{M} \\
&\Rightarrow (\exists t' \in \overline{C}) P(t)=P(t'), t'\sigma \in F(K) \\
&\Rightarrow D(\overline{t'\sigma}) \cap \overline{M} \subseteq \overline{K}\ \ \mbox{(by definition of $F(K)$)}.
\end{align*}
Then by (\ref{eq:D})
\begin{align*}
\bigcup \left\{ [ \overline{t'\sigma} \cap \overline{C}.\sigma ] \cap  \overline{C}.\sigma \ |\ \sigma \in \Sigma \right\} \cap \overline{M} \subseteq \overline{K}.
\end{align*}
Since $t\sigma$ belongs to the left-hand-side of the above inequality, we have $t\sigma \in \overline{K} = F(K)$.
Therefore $D(F(K)) \cap \overline{M} \subseteq F(K)$; equivalently $D(\overline{K}) \cap \overline{M} \subseteq \overline{K}$. This completes the proof of $K \in \mathcal {O}(C)$.  
\hfill $\square$

In view of Proposition~\ref{lem:fixpoint}, it is natural to attempt to compute $\sup \mathcal {O}(C)$ by iteration of $\Omega$ as follows:
\begin{align} \label{eq:Kj}
(\forall j \geq 1)\ K_j = \Omega(K_{j-1}),\ \ \ K_0 = C.
\end{align}
It is readily verified that $\Omega(K) \subseteq K$; hence
\begin{align*}
K_0 \supseteq K_1 \supseteq K_2 \supseteq \cdots
\end{align*}
Namely the sequence $\{K_j\}$ ($j \geq 1$)
is a monotone (descending) sequence of languages.
This implies that the (set-theoretic) limit 
\begin{align} \label{eq:Klimit}
K_\infty := \lim_{j \rightarrow \infty} K_j = \bigcap^{\infty}_{j=0} K_j
\end{align}
exists.  The following result asserts that if $K_\infty$ is reached in a
{\it finite} number of steps, then $K_\infty$ is precisely the supremal relatively
observable sublanguage of $C$, i.e. $\sup \mathcal {O}(C)$.

\begin{prop} \label{lem:limit}
If $K_\infty$ in (\ref{eq:Klimit}) is reached in a finite number of
steps, then
\begin{align*}
K_\infty = \sup \mathcal {O}(C).
\end{align*}
\end{prop}
\medskip

\emph{Proof.} Suppose that the limit $K_\infty$ is reached in a finite
number of steps. Then $K_\infty = \Omega(K_\infty)$. As in the proof of Proposition~\ref{lem:fixpoint}, we derive that $K_\infty \in \mathcal {O}(C)$.

It remains to show that $K_\infty$ is the supremal element of $\mathcal {O}(C)$. Let $K' \in \mathcal {O}(C)$; it will be shown that $K' \subseteq K_\infty$ by induction.
The base case $K' \subseteq K_0$ holds because $K' \subseteq C$ and $K_0 = C$. Suppose that $K' \subseteq K_{j-1}$. Let $s \in \overline{K'}$. Then
$s \in \overline{K_{j-1}}$ and
\begin{align*}
D(\overline{s}) \cap \overline{M} &\subseteq D(\overline{K'}) \cap \overline{M} \\
&\subseteq \overline{K'} \ \ \ \mbox{(by $K' \in \mathcal {O}(C)$)}\\
&\subseteq \overline{K_{j-1}}.
\end{align*}
Hence $s \in F(K_{j-1})$. This shows that 
\begin{align*}
& \overline{K'} \subseteq F(K_{j-1}) \\
& \Rightarrow K' \subseteq F(K_{j-1}) \\
& \Rightarrow K' \subseteq K_{j-1} \cap F(K_{j-1}).
\end{align*}
Moreover, since $K' \in \mathcal {O}(C)$, $K'$ is normal with respect to $\overline{C} \cap M$.
Thus $K' \subseteq \sup \mathcal {N} \big( K_{j-1} \cap F(K_{j-1}),\ \overline{C} \cap M \big) = K_{j}$. This completes the proof of the induction step, and therefore confirms that $K' \subseteq K_\infty$.
\hfill $\square$

In the next section, we shall establish that, when the given languages $M$ and $C$ are {\it regular}, the limit $K_\infty$ in (\ref{eq:Klimit}) is indeed reached in a finite number of steps.


\section{Effective Computation of $\sup \mathcal {O}(C)$ in the Regular Case} \label{sec3_reglan}


In this section, we first review the concept of Nerode equivalence relation and a finite convergence result for a sequence of regular languages. Based on these, we then prove that the sequence generated by (\ref{eq:Kj}) converges to the supremal relatively observable sublanguage $\sup \mathcal {O}(C)$ in a finite number of steps. Finally, we show that the computation of $\sup \mathcal {O}(C)$ is effective.

\subsection{Preliminaries} \label{subsec:preli}

Let $\pi$ be an arbitrary {\it equivalence relation} on $\Sigma^*$. Denote by $\Sigma^*/\pi$ the set of {\it equivalence classes} of $\pi$, and write $|\pi|$ for the cardinality of $\Sigma^*/\pi$. Define the {\it canonical projection} $P_\pi: \Sigma^* \rightarrow \Sigma^*/\pi$, namely the surjective function mapping any $s \in  \Sigma^*$ onto its equivalence class $P_\pi(s) \in \Sigma^*/\pi$.

Let $\pi_1, \pi_2$ be two equivalence relations on $\Sigma^*$. The {\it partial order} $\pi_1 \leq \pi_2$ holds if 
\begin{align*}
(\forall s_1,s_2 \in \Sigma^*)\ s_1 \equiv s_2 (\mbox{mod } \pi_1) \Rightarrow s_1 \equiv s_2 (\mbox{mod } \pi_2).
\end{align*}
The {\it meet} $\pi_1 \wedge \pi_2$ is defined by
\begin{align*}
(\forall s_1,s_2 \in \Sigma^*)\ s_1 \equiv s_2 (\mbox{mod } \pi_1 \wedge \pi_2) \mbox{ iff } s_1 \equiv s_2 (\mbox{mod } \pi_1) \ \&\  s_1 \equiv s_2 (\mbox{mod } \pi_2).
\end{align*}

For a language $L \subseteq \Sigma^*$, write $\mbox{Ner}(L)$ for the {\it Nerode
equivalence relation} \cite{HopUll:79} on $\Sigma^*$ with respect to $L$; namely for
all $s_1, s_2 \in \Sigma^*$, $s_1 \equiv s_2 (\mbox{mod }
\mbox{Ner}(L))$ provided
\begin{align*}
(\forall w \in \Sigma^*)\ s_1 w \in L \Leftrightarrow s_2 w \in L.
\end{align*}
Write $||L||$ for the cardinality of the set of equivalence classes
of $\mbox{Ner}(L)$, i.e. $||L||:=|\mbox{Ner}(L)|$. The language $L$
is said to be \emph{regular} \cite{HopUll:79} if $||L|| < \infty$. Henceforth, we assume that the given languages $M$ and $C$ are regular.

An equivalence relation $\rho$ is a {\it right congruence} on $\Sigma^*$ if 
\begin{align*}
(\forall s_1, s_2, t \in \Sigma^*)\  s_1 \equiv s_2 (\mbox{mod } \rho) \Rightarrow s_1 t \equiv s_2 t (\mbox{mod } \rho).
\end{align*}
Any Nerode equivalence relation is a right congruence.
For a right congruence $\rho$ and languages $L_1, L_2 \subseteq \Sigma^*$, we say that $L_1$ is {\it $\rho$-supported on} $L_2$ \cite[Section~2.8]{SCDES} if $\overline{L_1} \subseteq \overline{L_2}$ and 
\begin{align} \label{eq:support}
\{\overline{L_1}, \Sigma^*-\overline{L_1}\} \wedge \rho \wedge \mbox{Ner}(L_2) \leq \mbox{Ner}(L_1).
\end{align}
The $\rho$-support relation is {\it transitive}: namely, if $L_1$ is $\rho$-supported on $L_2$, and $L_2$ is $\rho$-supported on $L_3$, then $L_1$ is $\rho$-supported on $L_3$.
The following lemma is central to establish finite convergence of a monotone language sequence. 

\begin{lem} \label{lem:support} \cite[Theorem~2.8.11]{SCDES}
Given a monotone sequence of languages $K_0 \supseteq K_1 \supseteq K_2 \supseteq \cdots$ with $K_0$ regular, and a fixed right congruence $\rho$ on $\Sigma^*$ with $|\rho| < \infty$, suppose that $K_j$ is $\rho$-supported on $K_{j-1}$ for all $j \geq 1$. Then each $K_j$ is regular, and the sequence is finitely convergent to a sublanguage $K$. Furthermore, $K$ is supported on $K_0$ and
\begin{align*}
|| K || \leq |\rho| \cdot || K_0 || +1.
\end{align*}
\end{lem}

In view of this lemma, to show finite convergence of the sequence in (\ref{eq:Kj}), it suffices to find a fixed right congruence $\rho$ with $|\rho|<\infty$ such that $K_j$ is $\rho$-supported on $K_{j-1}$ for all $j \geq 1$. To this end, we need the following notation.

Let $\mu := \mbox{Ner}(M)$, $\eta := \mbox{Ner}(C)$ be Nerode equivalence relations and 
\begin{align*}
& \varphi_j := \{F(K_j), \Sigma^* - F(K_j)\},\ \kappa_j :=
\{\overline{K_j}, \Sigma^* - \overline{K_j}\}\ \ \ (j \geq 1)
\end{align*}
also stand for the equivalence relations corresponding to these partitions. 
Then $|\mu| <\infty$, $|\eta| <\infty$, and $|\varphi_j| = |\kappa_j| =2$.
Let $\pi$ be an equivalence relation on $\Sigma^*$, and define $f_\pi : \Sigma^* \rightarrow \mbox{Pwr}(\Sigma^*/\pi)$
according to
\begin{align} \label{eq:wp}
(\forall s \in \Sigma^*)\ f_\pi(s) = \{P_\pi(s') \ |\ s' \in [s] \cap \left( \overline{C}\cap M \right)\}
\end{align}
where $[s] = P^{-1} P (\{s\})$.
Write $\wp(\pi) := \ker\,f_\pi$. The size of $\wp(\pi)$
is $|\wp(\pi)| \leq 2^{|\pi|}$ \cite[Ex. 1.4.21]{SCDES}. Another property of
$\wp(\cdot)$ we shall use later is \cite[Ex. 1.4.21]{SCDES}:
\begin{align*}
\wp(\pi_1 \wedge \wp(\pi_2)) = \wp(\pi_1 \wedge \pi_2) = \wp(\wp(\pi_1) \wedge \pi_2)
\end{align*}
where $\pi_1, \pi_2$ are equivalence relations on $\Sigma^*$.

\subsection{Convergence Result}

First, we present a key result on support relation of the sequence $\{K_j\}$ generated by (\ref{eq:Kj}).

\begin{prop} \label{prop:support}
Consider the sequence $\{K_j\}$ generated by (\ref{eq:Kj}). For each $j \geq 1$, there holds that  $K_j$ is $\rho$-supported on $K_{j-1}$, where 
\begin{align} \label{eq:rho-support}
\rho := \mu \wedge \eta \wedge \wp( \mu \wedge \eta ).
\end{align}
\end{prop}

Let us postpone the proof of Proposition~\ref{prop:support}, and present immediately 
our main result.

\begin{theorem} \label{thm:convergence}
Consider the sequence $\{K_j\}$ generated by (\ref{eq:Kj}), and suppose that the given languages $M$ and $C$ are regular.  Then the sequence $\{K_j\}$ is finitely convergent
to $\sup \mathcal {O}(C)$, and $\sup \mathcal
{O}(C)$ is a regular language with
\begin{align*}
||\sup \mathcal {O}(C)|| \leq ||M|| \cdot ||C||
\cdot 2^{||M|| \cdot ||C||} + 1.
\end{align*}
\end{theorem}

{\it Proof.}  
Let $\rho = \mu \wedge \eta \wedge \wp(\mu \wedge \eta)$ as in (\ref{eq:rho-support}).
Since $\mu$ and $\eta$ are right congruences, so are $\mu \wedge \eta$ 
and $\wp(\mu \wedge \eta)$ (\cite[Example~6.1.25]{SCDES}). Hence $\rho$ is a right congruence, with \begin{align*}
|\rho| &\leq |\mu| \cdot |\eta| \cdot 2^{|\mu| \cdot |\eta|} \\
&= ||M|| \cdot ||C|| \cdot 2^{||M|| \cdot ||C||}.
\end{align*}
Since the languages $M$ and $C$ are regular, i.e. $||M||, ||C|| < \infty$, we derive that $|\rho| <\infty$.

It then follows from Lemmas~\ref{lem:limit} and \ref{lem:support} that the sequence $\{K_j\}$ is finitely convergent to $\sup \mathcal {O}(C)$, and $\sup \mathcal {O}(C)$ is $\rho$-supported on $K_0$, i.e.
\begin{align*}
\mbox{Ner}(\sup \mathcal {O}(C)) &\geq \{\overline{\sup \mathcal {O}(C)}, \Sigma^* - \overline{\sup \mathcal {O}(C)}\} \wedge \rho \wedge \mbox{Ner}(K_0) \\
&=  \{\overline{\sup \mathcal {O}(C)}, \Sigma^* - \overline{\sup \mathcal {O}(C)}\} \wedge \mu \wedge \eta \wedge \wp( \mu \wedge \eta ) \wedge \mbox{Ner}(K_0) \\
&=  \{\overline{\sup \mathcal {O}(C)}, \Sigma^* - \overline{\sup \mathcal {O}(C)}\} \wedge \mu \wedge \wp( \mu \wedge \eta ) \wedge \mbox{Ner}(K_0).
\end{align*}
Hence $\sup \mathcal {O}(C)$ is in fact ($\mu \wedge \wp( \mu \wedge \eta )$)-supported on $K_0$, which implies
\begin{align*}
||\sup \mathcal {O}(C)|| &\leq |\mu \wedge \wp( \mu \wedge \eta)| \cdot ||K_0|| +1 \\
&\leq ||M|| \cdot ||C||
\cdot 2^{||M|| \cdot ||C||} + 1 < \infty.
\end{align*}
Therefore $\sup \mathcal {O}(C)$ is itself a regular language.
\hfill $\square$

Theorem~\ref{thm:convergence} establishes the finite convergence of the sequence $\{K_j\}$ in (\ref{eq:Kj}), as well as the fact that an upper bound of $||\sup \mathcal {O}(C)||$ is
exponential in the product of $||M||$ and $||C||$.

In the sequel we prove Proposition~\ref{prop:support}, for which we need two lemmas.

\begin{lem} \label{Fj}
For each $j \geq 1$, the Nerode equivalence relation on $\Sigma^*$ with respect to $F(K_{j-1})$ satisfies
\begin{align*}
\mbox{Ner}(F(K_{j-1})) \geq \varphi_j \wedge \mbox{Ner}(K_{j-1}) \wedge
\wp(\mbox{Ner}(K_{j-1}) \wedge \mu \wedge \eta).
\end{align*}
\end{lem}

{\it Proof.}  First, let $s_1, s_2 \in \Sigma^* - F(K_{j-1})$; then for all $w \in \Sigma^*$ it holds that $s_1 w, s_2 w \in  \Sigma^* - F(K_{j-1})$. Thus $s_1 \equiv s_2 (\mbox{mod } \mbox{Ner}(F(K_{j-1})) )$.

Next, let $s_1, s_2 \in F(K_{j-1})$ and assume that 
\[
s_1 \equiv s_2 (\mbox{mod } \mbox{Ner}(K_{j-1}) \wedge
\wp(\mbox{Ner}(K_{j-1}) \wedge \mu \wedge \eta) ).
\] 
Also let $w \in \Sigma^*$ be such that $s_1 w \in F(K_{j-1})$. It will be shown that $s_2 w \in F(K_{j-1})$. Note first that $s_2 w \in \overline{K_{j-1}}$, since $s_1 w \in F(K_{j-1}) \subseteq \overline{K_{j-1}}$ and $s_1 \equiv s_2 (\mbox{mod } \mbox{Ner}(K_{j-1}) )$. Hence it is left to show that $D(\overline{s_2 w}) \cap \overline{M} \subseteq \overline{K_{j-1}}$, i.e.
\begin{align*}
\bigcup \left\{ [ \overline{s_2 w} \cap \overline{C}.\sigma ] \cap  \overline{C}.\sigma \ |\ \sigma \in \Sigma \right\} \cap \overline{M} \subseteq \overline{K_{j-1}}.
\end{align*}
It follows from $s_2 \in F(K_{j-1})$ that
\begin{align*}
\bigcup \left\{ [ \overline{s_2} \cap \overline{C}.\sigma ] \cap  \overline{C}.\sigma \ |\ \sigma \in \Sigma \right\} \cap \overline{M} \subseteq \overline{K_{j-1}}.
\end{align*}
Thus let $s'_2 \in [s_2]$, $x' \in [\overline{w}]$, and $s'_2 x' \in [ \overline{s_2 w} \cap \overline{C}.\sigma ] \cap  \overline{C}.\sigma \cap \overline{M}$ for some $\sigma \in \Sigma$. Write $x' := y' \sigma$, $y' \in \Sigma^*$. 
Since $s_1 \equiv s_2 (\mbox{mod } \wp(\mbox{Ner}(K_{j-1}) \wedge \mu \wedge \eta) )$, there exists $s'_1 \in [s_1]$ such that $s'_1 \equiv s'_2 (\mbox{mod } \mbox{Ner}(K_{j-1}) \wedge \mu \wedge \eta)$. Hence $s'_1 x' \in \overline{M}$ and $s'_1 y' \in \overline{C}$, and we derive that 
$s'_1 x' = s'_1 y' \sigma \in [ \overline{\{s_1 w\}} \cap \overline{C}.\sigma ] \cap  \overline{C}.\sigma \cap \overline{M}$. It then follows from $s_1 w \in F(K_{j-1})$ that $s'_1 x' \in \overline{K_{j-1}}$, which in turn implies that $s'_2 x' \in \overline{K_{j-1}}$. This completes the proof of $s_2 w \in F(K_{j-1})$, as required. \hfill $\square$

\begin{lem} \label{Kj}
For $K_j$ ($j \geq 1$) generated by (\ref{eq:Kj}), the following statements hold:
\begin{align*}
&K_{j} = \bigcup \left\{ [s] \cap \left( \overline{C} \cap M \right) \ |\ s \in \Sigma^* \ \&\ [s] \cap \left( \overline{C} \cap M \right)  \subseteq K_{j-1} \cap F(K_{j-1}) \right\}; \\
&\mbox{Ner}(K_{j}) \geq \mu \wedge \eta \wedge \wp(\mbox{Ner}(K_{j-1}) \wedge \mbox{Ner}(F(K_{j-1})) \wedge \mu \wedge \eta).
\end{align*}
\end{lem}

{\it Proof.} By (\ref{eq:Omega}) we know that $K_j$ is the supremal normal sublanguage of $K_{j-1} \cap F(K_{j-1})$ with respect to $\overline{C} \cap M$. Thus the conclusions follow immediately from Example~6.1.25 of \cite{SCDES}. \hfill $\square$

Now we are ready to prove Proposition~\ref{prop:support}.

\emph{Proof of Proposition~\ref{prop:support}.} 
To prove that $K_j$ is $\rho$-supported on $K_{j-1}$ ($j \geq 1$), by definition we must show that 
\begin{align*}
\mbox{Ner}(K_{j}) \geq \kappa_{j} \wedge \mu \wedge \eta \wedge \wp(\mu \wedge \eta) \wedge \mbox{Ner}(K_{j-1}).
\end{align*}
It suffices to show the following:
\begin{align*}
\mbox{Ner}(K_{j}) \geq \kappa_{j} \wedge \mu \wedge \eta \wedge \wp(\mu \wedge \eta).
\end{align*}
We prove this statement
by induction. First, we show the base case ($j=1$)
\begin{align*}
\mbox{Ner}(K_{1}) \geq \kappa_1 \wedge \mu \wedge \eta \wedge \wp(\mu \wedge \eta).
\end{align*}
From Lemma~\ref{Fj} and $K_0 = C$ (thus $\mbox{Ner}(K_{0}) = \eta$) we have
\begin{align*}
\mbox{Ner}(F(K_0)) &\geq \varphi_1 \wedge \mbox{Ner}(K_{0}) \wedge
\wp(\mbox{Ner}(K_{0}) \wedge \mu \wedge \eta) \\
&= \varphi_1 \wedge \eta \wedge \wp(\mu \wedge \eta).
\end{align*}
It then follows from Lemma~\ref{Kj} that
\begin{align}
\mbox{Ner}(K_{1}) &\geq \mu \wedge \eta \wedge \wp(\mbox{Ner}(K_{0}) \wedge \mbox{Ner}(F(K_0)) \wedge \mu \wedge \eta) \notag\\
&\geq \mu \wedge \eta \wedge \wp(\eta \wedge \varphi_1 \wedge \eta \wedge \wp(\mu \wedge \eta) \wedge \mu \wedge \eta) \notag\\
&= \mu \wedge \eta \wedge \wp(\varphi_1 \wedge \mu \wedge \eta) \wedge \wp(\mu \wedge \eta) \notag\\
&= \mu \wedge \eta \wedge \wp(\varphi_1 \wedge \mu \wedge \eta). \label{eq:base}
\end{align}
We claim that
\[
\mbox{Ner}(K_{1}) \geq \kappa_1 \wedge \mu \wedge \eta \wedge \wp(\mu \wedge \eta).
\] 
To show this, let $s_1, s_2
\in \Sigma^*$ and assume that $s_1 \equiv s_2 (\mbox{mod }
\kappa_{1} \wedge \mu \wedge \eta \wedge \wp(\mu \wedge \eta))$. If
$s_1, s_2 \in \Sigma^*-\overline{K_{1}}$, then for all $w \in \Sigma^*$, $s_1 w, s_2 w \in \Sigma^*-\overline{K_{1}}$; thus $s_1 \equiv s_2 (\mbox{mod } \mbox{Ner}(K_{1}))$. Now let $s_1, s_2 \in
\overline{K_{1}}$. By Lemma~\ref{Kj} we derive that for all $s'_1
\in [s_1] \cap \left( \overline{C} \cap M \right)$ and $s'_2 \in [s_2] \cap \left( \overline{C} \cap M \right)$, $s'_1, s'_2 \in
\overline{K_{1}}$. Since $\overline{K_{1}} \subseteq F(K_0)$, $s'_1, s'_2 \in F(K_0)$ and hence
\begin{align*}
\{P_{\varphi_1 \wedge \mu \wedge \eta}(s_1') \ |\ s_1'
\in [s_1] \cap \left( \overline{C} \cap M \right)\} = \{P_{\varphi_1 \wedge \mu \wedge \eta}(s_2') \ |\ s_2' \in [s_2] \cap \left( \overline{C} \cap M \right)\}.
\end{align*}
Namely $s_1 \equiv s_2 (\mbox{mod } \wp(\varphi_1 \wedge \mu \wedge \eta))$. This implies that $s_1 \equiv s_2 (\mbox{mod } \mbox{Ner}(K_{1}))$ by (\ref{eq:base}). Hence the above claim is established, and the base case is proved.

For the induction step, suppose that for $j \geq 2$, there holds
\begin{align*}
\mbox{Ner}(K_{j-1}) \geq \kappa_{j-1} \wedge \mu \wedge \eta \wedge \wp(\mu \wedge \eta).
\end{align*}
Again by Lemma~\ref{Fj} we have
\begin{align*}
\mbox{Ner}(F(K_{j-1})) &\geq \varphi_{j-1} \wedge \mbox{Ner}(K_{j-1}) \wedge
\wp(\mbox{Ner}(K_{j-1}) \wedge \mu \wedge \eta) \\
&\geq \varphi_{j-1} \wedge \kappa_{j-1} \wedge \mu \wedge \eta \wedge \wp(\mu \wedge \eta) \wedge \wp(\kappa_{j-1} \wedge \mu \wedge \eta \wedge \wp(\mu \wedge \eta) \wedge \mu \wedge \eta) \\
&= \varphi_{j-1} \wedge \kappa_{j-1} \wedge \mu \wedge \eta \wedge \wp(\mu \wedge \eta) \wedge \wp(\kappa_{j-1} \wedge \mu \wedge \eta) \\
&= \varphi_{j-1} \wedge \kappa_{j-1} \wedge \mu \wedge \eta \wedge \wp(\kappa_{j-1} \wedge \mu \wedge \eta) 
\end{align*}
Then by Lemma~\ref{Kj},
\begin{align}
\mbox{Ner}(K_{j}) &\geq \mu \wedge \eta \wedge \wp(\mbox{Ner}(K_{j-1}) \wedge \mbox{Ner}(F(K_{j-1})) \wedge \mu \wedge \eta) \notag\\
&\geq \mu \wedge \eta \wedge \wp(\varphi_{j-1} \wedge \kappa_{j-1} \wedge \mu \wedge \eta \wedge
\wp(\kappa_{j-1} \wedge \mu \wedge \eta)) \notag\\
&= \mu \wedge \eta \wedge \wp(\varphi_{j-1} \wedge \kappa_{j-1} \wedge \mu \wedge \eta). \label{eq:induction}
\end{align}
We claim that 
\[ 
\mbox{Ner}(K_{j}) \geq \kappa_{j} \wedge \mu \wedge \eta
\wedge \wp(\mu \wedge \eta). 
\]
To show this, let $s_1, s_2 \in \Sigma^*$ and assume that $s_1 \equiv s_2 (\mbox{mod }
\kappa_{j} \wedge \mu \wedge \eta \wedge \wp(\mu \wedge \eta))$. If
$s_1, s_2 \in \Sigma^*-\overline{K_{j}}$, then for all $w \in \Sigma^*$, $s_1 w, s_2 w \in \Sigma^*-\overline{K_{j}}$; hence $s_1 \equiv s_2
(\mbox{mod } \mbox{Ner}(K_{j}))$. Now let $s_1, s_2 \in
\overline{K_{j}}$. By Lemma~\ref{Kj} we derive that for all $s'_1
\in [s_1] \cap \left( \overline{C} \cap M \right)$ and $s'_2 \in [s_2] \cap \left( \overline{C} \cap M \right)$, $s'_1, s'_2 \in \overline{K_{j}}$. Since $\overline{K_{j}} \subseteq F(K_{j-1})
\subseteq \overline{K_{j-1}}$,
\begin{align*}
&\{P_{\varphi_{j-1} \wedge \kappa_{j-1} \wedge \mu \wedge \eta}(s_1') \ |\ s_1'
\in [s_1] \cap \left( \overline{C} \cap M \right)\} \\
= &\{P_{\varphi_{j-1} \wedge \kappa_{j-1} \wedge \mu \wedge \eta}(s_2') \ |\ s_2' \in [s_2] \cap \left( \overline{C} \cap M \right)\}.
\end{align*}
Namely $s_1 \equiv s_2 (\mbox{mod } \wp(\varphi_{j-1} \wedge \kappa_{j-1}
\wedge \mu \wedge \eta)$. This implies that $s_1 \equiv
s_2 (\mbox{mod } \mbox{Ner}(K_{j}))$ by (\ref{eq:induction}). Therefore the above claim
is established, and the induction step is completed. 

\hfill $\square$

\subsection{Effective Computability of $\Omega$} \label{subsec:effcom}

We conclude this section by showing that the iteration scheme in (\ref{eq:Kj}) yields an effective procedure for the computation of $\sup \mathcal {O}(C)$, when the given languages $M$ and $C$ are regular.  For this, owing to Theorem~\ref{thm:convergence}, it suffices to prove that the operator $\Omega$ in (\ref{eq:Omega}) is effectively computable. 

Recall that a language $L \subseteq \Sigma^*$ is regular if and only if there exists a finite-state automaton ${\bf G} = (Q, \Sigma, \delta, q_0, Q_m)$ such that 
\begin{align*}
L_m({\bf G}) = \{s \in \Sigma^* \ |\ \delta(q_0,s) \in Q_m\} = L.
\end{align*}
Let $\mathcal{O} : (Pwr(\Sigma^*))^k \rightarrow (Pwr(\Sigma^*))$ be an operator that preserves regularity; namely $L_1,...,L_k$ regular implies $\mathcal{O}(L_1,...,L_k)$ regular. We say that $\mathcal{O}$ is {\it effectively computable} if from each $k$-tuple $(L_1,...,L_k)$ of regular languages, one can construct a finite-state automaton ${\bf G}$ with $L_m({\bf G}) = \mathcal{O}(L_1,...,L_k)$.

The standard operators of language closure, complement,\footnote{For a language $L \subseteq \Sigma^*$, its complement, written $L^c$, is $\Sigma^* - L$.} union, and intersection all preserve regularity and are effectively computable \cite{Eil:74}. Moreover, both the operator $\sup \mathcal {N} : Pwr(\Sigma^*) \rightarrow Pwr(\Sigma^*)$ given by 
\begin{align*}
\sup \mathcal {N}(L) := \bigcup \{ L' \subseteq L \ |\ [L'] \cap H = L' \}, \ \ \ \mbox{for some fixed } H \subseteq \Sigma^*
\end{align*}
and the operator $\sup \mathcal {F} : Pwr(\Sigma^*) \rightarrow Pwr(\Sigma^*)$ given by 
\begin{align*}
\sup \mathcal {F}(L) := \bigcup \{ L' \subseteq L \ |\ \overline{L'} = L' \}
\end{align*}
preserve regularity and are effectively computable (see \cite{ChoMarcus:89} and \cite{WonRam:87}, respectively).

The main result of this subsection is the following theorem.

\begin{theorem} \label{thm:effcom}
Suppose that $M$ and $C$ are regular. Then
the operator $\Omega$ in (\ref{eq:Omega}) preserves regularity and is effectively computable.
\end{theorem}

The following proposition is a key fact.

\begin{prop} \label{eq:FK_effcom}
For each $K \subseteq \Sigma^*$,
\begin{align*}
F(K) = \overline{K} \cap \sup \mathcal {F} \left( \bigcap\{ \sup \mathcal {N}(\overline{K} \cup (\overline{M} \cap \overline{C}.\sigma)^c) \cup (\overline{C}.\sigma)^c \ |\ \sigma \in \Sigma \} \right).
\end{align*}
\end{prop}

{\it Proof.} 
By (\ref{eq:FK}) and (\ref{eq:D}), 
\begin{align*}
F(K) = \{ s \in \overline{K} \ |\ \bigcup \left\{ [ \overline{s} \cap
\overline{C}.\sigma ] \cap  \overline{C}.\sigma \ |\ \sigma \in \Sigma \right\} \cap \overline{M} \subseteq \overline{K} \}.
\end{align*}
Hence
\begin{align*}
s \in F(K) &\Leftrightarrow s \in \overline{K} \mbox{ and } \bigcup \left\{ [ \overline{s} \cap
\overline{C}.\sigma ] \cap  \overline{C}.\sigma \ |\ \sigma \in \Sigma \right\} \cap \overline{M} \subseteq \overline{K} \\
&\Leftrightarrow s \in \overline{K} \mbox{ and } \bigcup \left\{ [ \overline{s} \cap
\overline{C}.\sigma ] \cap  \overline{C}.\sigma \ |\ \sigma \in \Sigma \right\} \subseteq \overline{K} \cup (\overline{M})^c \\
&\Leftrightarrow s \in \overline{K} \mbox{ and } (\forall \sigma \in \Sigma)\ [ \overline{s} \cap
\overline{C}.\sigma ] \cap  \overline{C}.\sigma \subseteq \overline{K} \cup (\overline{M})^c  \\
&\Leftrightarrow s \in \overline{K} \mbox{ and } (\forall \sigma \in \Sigma)\ [ \overline{s} \cap
\overline{C}.\sigma ] \subseteq \overline{K} \cup (\overline{M})^c \cup (\overline{C}.\sigma)^c  \\
&\Leftrightarrow s \in \overline{K} \mbox{ and } (\forall \sigma \in \Sigma)\ [ \overline{s} \cap
\overline{C}.\sigma ] \subseteq \overline{K} \cup (\overline{M} \cap \overline{C}.\sigma)^c  \\
&\Leftrightarrow s \in \overline{K} \mbox{ and } (\forall \sigma \in \Sigma)\  \overline{s} \cap
\overline{C}.\sigma \subseteq \sup \mathcal{N} (\overline{K} \cup (\overline{M} \cap \overline{C}.\sigma)^c)  \\
&\Leftrightarrow s \in \overline{K} \mbox{ and } (\forall \sigma \in \Sigma)\  \overline{s} \subseteq \sup \mathcal{N} (\overline{K} \cup (\overline{M} \cap \overline{C}.\sigma)^c) \cup (\overline{C}.\sigma)^c \\
&\Leftrightarrow s \in \overline{K} \mbox{ and } \overline{s} \subseteq \bigcap \{ \sup \mathcal{N} (\overline{K} \cup (\overline{M} \cap \overline{C}.\sigma)^c) \cup (\overline{C}.\sigma)^c \ |\ \sigma \in \Sigma \}\\
&\Leftrightarrow s \in \overline{K} \mbox{ and } s \in \sup \mathcal{F} \left( \bigcap \{ \sup \mathcal{N} (\overline{K} \cup (\overline{M} \cap \overline{C}.\sigma)^c) \cup (\overline{C}.\sigma)^c \ |\ \sigma \in \Sigma \} \right) \\
&\Leftrightarrow s \in \overline{K} \cap \sup \mathcal{F} \left( \bigcap \{ \sup \mathcal{N} (\overline{K} \cup (\overline{M} \cap \overline{C}.\sigma)^c) \cup (\overline{C}.\sigma)^c \ |\ \sigma \in \Sigma \} \right). 
\end{align*}
\hfill $\square$

We also need the following lemma.

\begin{lem} \label{lem:Csigma_effcom}
Let $\sigma \in \Sigma$ be fixed. Then the operator $B_\sigma : Pwr(\Sigma^*) \rightarrow Pwr(\Sigma^*)$ given by 
\begin{align*}
B_\sigma(L) := \overline{L}.\sigma = \{ s\sigma \ |\ s \in \overline{L} \}
\end{align*}
preserves regularity and is effectively computable.
\end{lem}

{\it Proof.} Let ${\bf G} = (Q, \Sigma, \delta, q_0, Q_m)$ be a finite-state automaton with $L_m({\bf G}) = L$. We will construct a new finite-state automaton {\bf H} such that $L_m({\bf H}) = B_\sigma(L)$. The construction is in two steps. First, let $q^*$ be a new state (i.e. $q^* \notin Q$), and define ${\bf G}' = (Q', \Sigma, \delta', q_0, Q'_m)$ where
\begin{align*}
Q' := Q \cup \{q^*\},\ \ \ \delta' := \delta \cup \{(q,\sigma,q^*) | q \in Q\},\ \ \ Q'_m := \{q^*\}.
\end{align*}
Thus ${\bf G}'$ is a finite-state automaton with $L_m({\bf G}') = B_\sigma(L)$. However, ${\bf G}'$ is {\it nondeterministic}, inasmuch as $\delta'(q,\sigma)=\{q',q^*\}$ whenever $\delta(q,\sigma)$ is defined and $\delta(q,\sigma)=q'$. The second step is hence to apply the standard subset construction to convert the nondeterministic ${\bf G}'$ to a {\it deterministic} finite-state automaton ${\bf H}$ with $L_m({\bf H}) =L_m({\bf G}')= B_\sigma(L)$. This completes the proof. \hfill $\square$

Finally we present the proof of Theorem~\ref{thm:effcom}.

{\it Proof of Theorem~\ref{thm:effcom}.} By Proposition~\ref{eq:FK_effcom} and the definition of $\Omega : Pwr(\Sigma^*) \rightarrow Pwr(\Sigma^*)$ in (\ref{eq:Omega}), for each $K \subseteq \Sigma^*$ we derive 
\begin{align*}
\Omega(K) = \sup \mathcal{N} \left( K \cap \sup \mathcal {F} \left( \bigcap\{ \sup \mathcal {N}(\overline{K} \cup (\overline{M} \cap \overline{C}.\sigma)^c) \cup (\overline{C}.\sigma)^c \ |\ \sigma \in \Sigma \} \right) \right).
\end{align*}
Since the language closure, complement, union, intersection, $\sup \mathcal{N}$, $\sup \mathcal{F}$ and $\overline{C}.\sigma$ (by Lemma~\ref{lem:Csigma_effcom}) all preserve regularity and are effectively computable, the same conclusion for the operator $\Omega$ follows immediately.  \hfill $\square$

In the proof, we see that the operator $\Omega$ in (\ref{eq:Omega}) is decomposed into a sequence of standard or well-known language operations. This allows straightforward implementation of $\Omega$ using off-the-shelf algorithms.


\section{Relative Observability and Controllability} \label{sec4_obscon}

For the purpose of supervisory control under partial observation, we combine relative observability with {\it controllability} and provide a fixpoint characterization of the supremal relatively observable and controllable sublanguage.

Let the alphabet $\Sigma$ be partitioned into $\Sigma_c$, the subset of controllable events, and $\Sigma_u$, the subset of uncontrollable events. For the given $M$ and $C$, we say that $C$ is controllable with respect to $M$ if 
\begin{align*}
\overline{C}\Sigma_u \cap \overline{M} \subseteq \overline{C}.
\end{align*}
Whether or not $C$ is controllable, write $\mathcal{C}(C)$ for the family of all controllable sublanguages of $C$. Then the supremal element $\sup \mathcal{C}(C)$ exists and is effectively computable \cite{WonRam:87}.

Now write $\mathcal{CO}(C)$ for the family of controllable and $C$-observable sublanguages of $C$. Note that the family $\mathcal{CO}(C)$ is nonempty inasmuch as the empty language is a member.  Thanks to the closed-under-union property of both controllability and $C$-observability, the supremal controllable and $C$-observable sublanguage $\sup \mathcal{CO}(C)$ therefore exists and is given by
\begin{align} \label{eq:supCO}
\sup \mathcal {CO}(C) := \bigcup \{ K \ |\ K \in \mathcal {CO}(C) \}.
\end{align}

Define the operator $\Gamma : Pwr(\Sigma^*) \rightarrow Pwr(\Sigma^*)$ by
\begin{align} \label{eq:Gamma}
\Gamma(K) := \sup \mathcal{O}( \sup \mathcal{C}(K) ).
\end{align}
The proposition below characterizes $\sup \mathcal{CO}(C)$ as the largest fixpoint of $\Gamma$.

\begin{prop} \label{lem:fixpointCO}
$\sup \mathcal {CO}(C) = \Gamma(\sup \mathcal {CO}(C))$, and $\sup
\mathcal {CO}(C) \supseteq K$ for every $K$ such that $K =
\Gamma(K)$.
\end{prop}

\emph{Proof.} Since $\sup \mathcal {CO}(C) \in \mathcal {CO}(C)$, i.e. both controllable and $C$-observable,
\begin{align*}
\Gamma(\sup \mathcal {CO}(C)) &= \sup \mathcal{O}( \sup \mathcal{C}(\sup \mathcal {CO}(C)) )\\
&= \sup \mathcal{O}( \sup \mathcal {CO}(C) )\\
&= \sup \mathcal {CO}(C).
\end{align*}

Next let $K$ be such that $K = \Gamma(K)$. To show that $K \subseteq \sup \mathcal {CO}(C)$, 
it suffices to show that $K \in \mathcal {CO}(C)$. Let $H := \sup \mathcal {C}(K)$; thus $H \subseteq K$.  On the other hand, from $K = \Gamma(K) = \sup \mathcal{O}( H )$ we have $K \subseteq H$. Hence $K = H$. It follows that $K =  \sup \mathcal {C}(K)$ and $K = \sup \mathcal{O}(K)$, which means that $K$ is both controllable and $C$-observable. Therefore we conclude that $K \in \mathcal {CO}(C)$.
\hfill $\square$

In view of Proposition~\ref{lem:fixpointCO}, we compute $\sup \mathcal {CO}(C)$ by iteration of $\Gamma$ as follows:
\begin{align} \label{eq:KjCO}
(\forall j \geq 1)\ K_j = \Gamma(K_{j-1}),\ \ \ K_0 = C.
\end{align}
It is readily verified that $\Gamma(K) \subseteq K$, and thus
\begin{align*}
K_0 \supseteq K_1 \supseteq K_2 \supseteq \cdots
\end{align*}
Namely the sequence $\{K_j\}$ ($j \geq 1$)
is a monotone (descending) sequence of languages.
Recalling the notation from Section~\ref{subsec:preli}, we have the following key result.

\begin{prop} \label{prop:supportCO}
Consider the sequence $\{K_j\}$ generated by (\ref{eq:KjCO}) and let $\rho = \mu \wedge \eta \wedge \wp( \mu \wedge \eta )$.  Then for each $j \geq 1$, $K_j$ is $\rho$-supported on $K_{j-1}$.
\end{prop}

{\it Proof.}  Write $H_j := \sup \mathcal {C}(K_{j-1})$ and $\psi_j := \{\overline{H_j}, \Sigma^*-\overline{H_j}\}$ for $j \geq 1$. Then by \cite[p.~642]{WonRam:87} there holds
\begin{align*}
\mbox{Ner}(H_{j}) \geq \psi_{j} \wedge \mu \wedge \mbox{Ner}(K_{j-1}).
\end{align*}
We claim that for $j \geq 1$,
\begin{align*}
\mbox{Ner}(K_{j}) \geq \kappa_{j} \wedge \mu \wedge \eta \wedge \wp(\mu \wedge \eta).
\end{align*}
We prove this claim by induction. For the base case ($j=1$), 
\begin{align*}
\mbox{Ner}(H_{1}) &\geq \psi_{1} \wedge \mu \wedge \mbox{Ner}(K_{0}) \\
&= \psi_{1} \wedge \mu \wedge \eta
\end{align*}
Since $K_1 = \sup \mathcal{O}(H_1)$, we set up the following sequence to compute $K_1$:
\begin{align*}
(\forall i \geq 1)\ T_i = \Omega(T_{i-1}),\ \ \ T_0 = H_1.
\end{align*}
Following the derivations in the proof of Proposition~\ref{prop:support}, it is readily shown that each $T_i$ is $\rho$-supported on $H_1$; in particular, 
\begin{align*}
\mbox{Ner}(K_{1}) &\geq \kappa_{1} \wedge \rho \wedge \mbox{Ner}(H_{1}) \\
&\geq  \kappa_{1} \wedge \psi_1 \wedge \mu \wedge \eta \wedge \wp(\mu \wedge \eta) \\
&=  \kappa_{1} \wedge \mu \wedge \eta \wedge \wp(\mu \wedge \eta).
\end{align*}
This confirms the base case.

For the induction step, suppose that for $j \geq 2$, there holds
\begin{align*}
\mbox{Ner}(K_{j-1}) \geq \kappa_{j-1} \wedge \mu \wedge \eta \wedge \wp(\mu \wedge \eta).
\end{align*}
Thus 
\begin{align*}
\mbox{Ner}(H_{j}) &\geq \psi_{j} \wedge \mu \wedge \mbox{Ner}(K_{j-1}) \\
&\geq \psi_{j} \wedge \kappa_{j-1} \wedge \mu \wedge \eta \wedge \wp(\mu \wedge \eta) \\
&= \psi_{j} \wedge \mu \wedge \eta \wedge \wp(\mu \wedge \eta).
\end{align*}
Again  set up a sequence to compute $K_j$ as follows:
\begin{align*}
(\forall i \geq 1)\ T_i = \Omega(T_{i-1}),\ \ \ T_0 = H_j.
\end{align*}
We derive by similar calculations as in Proposition~\ref{prop:support} that each $T_i$ is $\rho$-supported on $H_j$; in particular, 
\begin{align*}
\mbox{Ner}(K_{j}) &\geq \kappa_{j} \wedge \rho \wedge \mbox{Ner}(H_{j}) \\
&\geq \kappa_{j} \wedge \psi_j \wedge \mu \wedge \eta \wedge \wp(\mu \wedge \eta) \\
&=  \kappa_{j} \wedge \mu \wedge \eta \wedge \wp(\mu \wedge \eta).
\end{align*}
Therefore the induction step is completed, and the above claim
is established.  Then it follows immediately
\begin{align*}
\mbox{Ner}(K_{j}) &\geq \kappa_{j} \wedge \mu \wedge \eta \wedge \wp(\mu \wedge \eta) \wedge \mbox{Ner}(K_{j-1}) \\
&= \kappa_{j} \wedge \rho \wedge \mbox{Ner}(K_{j-1}).
\end{align*}
Namely, $K_j$ is $\rho$-supported on $K_{j-1}$, as required.
\hfill $\square$

The following theorem is the main result of this section, which follows directly from Proposition~\ref{prop:supportCO} and Lemma~\ref{lem:support}.

\begin{theorem} \label{thm:convergenceCO}
Consider the sequence $\{K_j\}$ in (\ref{eq:KjCO}), and suppose that the given languages $M$ and $C$ are regular.  Then the sequence $\{K_j\}$ is finitely convergent
to $\sup \mathcal {CO}(C)$, and $\sup \mathcal
{CO}(C)$ is a regular language with
\begin{align*}
||\sup \mathcal {CO}(C)|| \leq ||M|| \cdot ||C||
\cdot 2^{||M|| \cdot ||C||} + 1.
\end{align*}
\end{theorem}

Finally, $\sup \mathcal {CO}(C)$ is effectively computable, inasmuch as the operators $\sup \mathcal {C}(\cdot)$ and $\sup \mathcal {O}(\cdot)$ are (see \cite{WonRam:87} and Theorem~\ref{thm:effcom}, respectively). In particular, the operator $\Gamma$ in (\ref{eq:Gamma}) is effectively computable.


\section{Examples} \label{sec5_examp}

\begin{figure}[!t]
  \centering
  \includegraphics[width=0.75\textwidth]{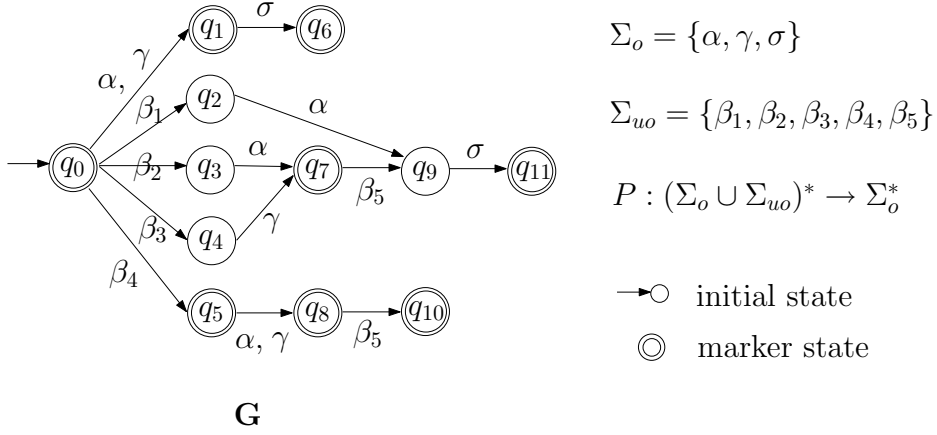}
  \caption{Example: computation of the supremal $C$-observable sublanguage $\sup \mathcal {O}(C)$
  by iteration of the operator $\Omega$ in (\ref{eq:Omega})}
  \label{fig:ex_supobs} 
\end{figure}

In this section, we first give an example to illustrate the computation of the supremal $C$-observable sublanguage $\sup \mathcal {O}(C)$ (by iteration of the operator $\Omega$). Then we present an empirical study on the computation of the supremal controllable and $C$-observable sublanguage $\sup \mathcal {CO}(C)$ (by iteration of the operator $\Gamma$, which has been implemented by a computer program).

\subsection{An Example of Computing $\sup \mathcal {O}(C)$}

Consider the example 
displayed in Fig.~\ref{fig:ex_supobs}. The observable event set is
$\Sigma_o = \{\alpha, \gamma, \sigma\}$ and unobservable
$\Sigma_{uo} = \{\beta_1,\beta_2,\beta_3,\beta_4,\beta_5\}$; thus
the natural projection is $P : (\Sigma_o \cup \Sigma_{uo})^*
\rightarrow \Sigma^*_o$. Let 
\begin{align*}
M := L_m({\bf G}) = \{ &\epsilon, \alpha, \gamma, \alpha \sigma, \gamma \sigma, \beta_1 \alpha \sigma,
\beta_2 \alpha,  \beta_2 \alpha \beta_5 \sigma, \beta_3 \gamma, \\
& \beta_3 \gamma \beta_5 \sigma, \beta_4, \beta_4 \alpha, \beta_4 \gamma, \beta_4 \alpha \beta_5, \beta_4 \gamma \beta_5  \}
\end{align*}
and the specification language
\begin{align*}
C := M - \{\beta_4 \alpha \beta_5, \beta_4 \gamma \beta_5\}.
\end{align*}
Both $M$ and $C$ are regular languages.

Now apply the operator $\Omega$ in (\ref{eq:Omega}). Initialize $K_0 = C$. The first iteration $j
= 1$ starts with
\begin{align*}
F(K_0) &= \{ s \in \overline{K_{0}} \ |\ D(\overline{s}) \cap \overline{M}
\subseteq \overline{K_{0}} \}\\
&= \{\epsilon, \alpha, \gamma, \alpha \sigma, \gamma \sigma, \beta_1, \beta_1 \alpha, \beta_1 \alpha \sigma, \beta_2, \beta_2 \alpha, \beta_3, \beta_3 \gamma, \beta_4, \beta_4 \alpha, \beta_4 \gamma\} \\
&= \overline{K_0} - \{\beta_2 \alpha \beta_5, \beta_2 \alpha \beta_5
\sigma, \beta_3 \gamma \beta_5, \beta_3 \gamma \beta_5 \sigma\}.
\end{align*}
Note that since $\beta_2 \alpha \beta_5 \sigma \in K_0$, strings
$\beta_2 \alpha \beta_5, \beta_2 \alpha \beta_5 \sigma \in
\overline{K_0}$. But $\beta_2 \alpha \beta_5, \beta_2 \alpha \beta_5
\sigma \notin F(K_0)$; this is because the string $\beta_4 \alpha
\beta_5$ belongs to $D(\overline{\beta_2 \alpha \beta_5}) \cap
\overline{M}$ and $D(\overline{\beta_2 \alpha \beta_5 \sigma})
\cap \overline{M}$, but $\beta_4 \alpha \beta_5$ does not belong to
$\overline{K_0}$. For the same reason, $\beta_3 \gamma \beta_5,
\beta_3 \gamma \beta_5 \sigma \in \overline{K_0}$ but $\beta_3
\gamma \beta_5, \beta_3 \gamma \beta_5 \sigma \notin F(K_0)$. Next calculate
\begin{align*}
F(K_0) \cap K_0
&= \{\epsilon, \alpha, \gamma, \alpha \sigma, \gamma \sigma, \beta_1 \alpha \sigma, \beta_2 \alpha, \beta_3 \gamma, \beta_4, \beta_4 \alpha, \beta_4 \gamma\} \\
&= K_0 - \{\beta_2 \alpha \beta_5 \sigma, \beta_3 \gamma \beta_5 \sigma\}.
\end{align*}
Removing strings $\beta_2 \alpha \beta_5 \sigma, \beta_3 \gamma
\beta_5 \sigma$ from $K_0$ makes $F(K_0) \cap K_0$ {\it not} normal with respect
to $\overline{C} \cap M$. Indeed, $\alpha \sigma, \beta_1
\alpha \sigma \in [\beta_2 \alpha \beta_5 \sigma] \cap \overline{C}
\cap M$ and $\gamma \sigma \in [ \beta_3 \gamma \beta_5
\sigma] \cap \overline{C} \cap M$ violate the normality
condition and therefore must also be removed. Hence,
\begin{align*}
K_{1} &= \sup \mathcal {N}(F(K_0) \cap K_0, \overline{C} \cap L_m({\bf G}))\\
&= \{\epsilon, \alpha, \gamma, \beta_2 \alpha, \beta_3 \gamma,
\beta_4, \beta_4 \alpha, \beta_4 \gamma\} \\
&= (F(K_0) \cap K_0) - \{\alpha \sigma, \beta_1 \alpha \sigma, \gamma \sigma\}.
\end{align*}
This completes the first iteration $j=1$.

Since $K_1 \subsetneqq K_0$, we proceed to $j=2$,
\begin{align*}
F(K_1) &= \{ s \in \overline{K_{1}} \ |\ D(\overline{s}) \cap \overline{M}
\subseteq \overline{K_{1}} \}\\
&= \{\epsilon, \gamma, \beta_2, \beta_3, \beta_3 \gamma, \beta_4,
\beta_4 \gamma\} \\
&= \overline{K_1} - \{\alpha, \beta_2 \alpha, \beta_4 \alpha\}.
\end{align*}
We see that $\alpha, \beta_2 \alpha, \beta_4 \alpha \in
\overline{K_1}$ but $\alpha, \beta_2 \alpha, \beta_4 \alpha \notin
F(K_1)$. This is because the string $\beta_1 \alpha \in
D(\overline{\alpha}) \cap \overline{M}$, $D(\overline{\beta_2
\alpha}) \cap \overline{M}$, and $D(\overline{\beta_4 \alpha})
\cap \overline{M}$, but $\beta_1 \alpha \notin \overline{K_1}$. Note
that $\beta_1 \alpha$ was in $\overline{K_0}$ since $\beta_1 \alpha
\sigma \in K_0$, but $\beta_1 \alpha \sigma$ was removed so as to
ensure normality of $K_1$; this in turn removed $\beta_1 \alpha$,
which now causes removal of strings $\alpha, \beta_2 \alpha, \beta_4
\alpha$ altogether. Continuing,
\begin{align*}
F(K_1) \cap K_1
&= \{\epsilon, \gamma, \beta_3 \gamma, \beta_4, \beta_4 \gamma\} \\
&= K_1 - \{\alpha, \beta_2 \alpha, \beta_4 \alpha\}.
\end{align*}
Removing strings $\alpha, \beta_2 \alpha, \beta_4 \alpha$ does not destroy normality of $K_1$. Indeed $F(K_1) \cap K_1$ is normal with respect to $\overline{C} \cap M$ and we have
\begin{align*}
K_{2} &= \sup \mathcal {N}(F(K_1) \cap K_1, \overline{C} \cap M)\\
&= \{\epsilon, \gamma, \beta_3 \gamma, \beta_4, \beta_4 \gamma\}\\
&= F(K_1) \cap K_1.
\end{align*}
This completes the second iteration $j=2$.

Since $K_2 \subsetneqq K_1$, we proceed to $j=3$ as follows:
\begin{align*}
F(K_2) &= \{ s \in \overline{K_{2}} \ |\ D(\overline{s}) \cap \overline{M}
\subseteq \overline{K_{2}} \}\\
&= \{\epsilon, \gamma, \beta_3, \beta_3 \gamma, \beta_4, \beta_4 \gamma\} = \overline{K_2}; \\
F(K_2) \cap K_2 &= \overline{K_2} \cap K_2 = K_2;\\
K_{3} &= \sup \mathcal {N}(F(K_2) \cap K_2, \overline{C} \cap M)\\
&= \sup \mathcal {N}(K_2, \overline{C} \cap M) =K_2.
\end{align*}
Since $K_3 = K_2$, the limit of the sequence in (\ref{eq:Kj}) is reached. Therefore
\begin{align*}
K_3 = \{\epsilon, \gamma, \beta_3 \gamma, \beta_4, \beta_4 \gamma\}
\end{align*}
is the supremal $C$-observable sublanguage of $C$.

\subsection{A Case Study of Computing $\sup \mathcal {CO}(C)$} \label{subsec:agv}

Consider the same case study as in \cite[Section~V-B]{CaiZhaWon_TAC14}, namely a manufacturing workcell served by five automated guided vehicles (AGV). Adopting the same settings, we apply the implemented $\Gamma$ operator to compute the supremal relatively observable and controllable sublanguage $\sup \mathcal {CO}(C)$, as represented by a finite-state automaton, say {\bf SUPO}. That is, 

\noindent $L_m({\bf SUPO}) = \sup \mathcal {CO}(C)$.

For this case study, the full-observation supervisor (representing the supremal controllable sublanguage) has 4406 states and 11338 transitions. Selecting different subsets of unobservable events, the computational results for the supremal relatively observable and controllable sublanguages, or {\bf SUPO}, are listed in Table~\ref{tab:agv}. We see in all cases but the first ($\Sigma_{uo} = \{13\}$) that the state and transition numbers of {\bf SUPO} are fewer than those of the full-observation supervisor. When $\Sigma_{uo} = \{13\}$, in fact, the supremal controllable sublanguage is already observable, and is therefore itself the supremal relatively observable and controllable sublanguage.

Moreover, we have confirmed that the computation results agree with those by the algorithm in \cite{CaiZhaWon_TAC14}. Thus the new computation scheme provides a useful alternative to ensure presumed correctness based on consistency.

\begin{table*}[!t]
\renewcommand{\arraystretch}{1.3}
\caption{{\bf SUPO} computed for different subsets of
unobservable events in the AGV case study} \label{tab:agv}
\centering
\begin{tabular}{|c|c|}\hline
$\Sigma_{uo}=\Sigma-\Sigma_o$ & State \#, transition \# of {\bf SUPO} \\
\hline
\{13\} & (4406,11338)  \\
\hline
\{21\} & (4348,10810)  \\
\hline
\{31\} & (4302,11040)  \\
\hline
\{43\} & (4319,10923)  \\
\hline
\{51\} & (4400,11296)  \\
\hline
\{12,31\} & (1736,4440)  \\
\hline
\{24,41\} & (4122,10311)  \\
\hline
\{31,43\} & (4215,10639)  \\
\hline
\{32,51\} & (2692,6596)  \\
\hline
\{41,51\} & (3795,9355)  \\
\hline
\{11,31,41\} & (163,314)  \\
\hline
\{12,33,51\} & (94,140)  \\
\hline
\{12,24,33,44,53\} & (72,112)  \\
\hline
\{12,21,32,43,51\} & (166,314)  \\
\hline
\{13,23,31,33, & \multirow{2}{*}{(563,1244)} \\
41,43,51,53\} &  \\
\hline
\end{tabular}
\end{table*}


\section{Conclusions} \label{sec6_concl}

We have presented a new characterization of
relative observability, and an operator on languages whose largest fixpoint is the supremal relatively observable sublanguage. 
In the case of regular languages and based on the support relation, we have proved that the sequence of languages generated by the operator converges finitely to the supremal relatively observable sublanguage, and the operator is effectively computable. 

Moreover, for the purpose of supervisory control under partial observation, we have presented a second  operator that in the regular case effectively computes the supremal relatively observable and controllable sublanguage. Finally we have presented an example and a case study to illustrate the effectiveness of the proposed computation schemes.

%
%
\bibliographystyle{plain}
\bibliography{DES}

\end{document}